\documentclass[aps,pra,reprint,amsmath,amssymb]{revtex4-2}
\usepackage{graphicx}
\usepackage{color}
\usepackage{hyperref}
\usepackage{bm}
\usepackage{physics}
\usepackage{dsfont}
\usepackage{mathrsfs}

\newcommand{\eref}{Eq.~\eqref}

\begin{document}

\title{Markovian Embeddings of Non-Markovian Open System Dynamics}

\author{Meng Xu}
\email{meng.xu@uni-ulm.de}
\affiliation{
Institute  for Complex Quantum Systems and IQST, Ulm University, Albert-Einstein-Allee 11, 89069  Ulm, Germany}

\author{J\"urgen T. Stockburger}
\email{juergen.stockburger@uni-ulm.de}
\affiliation{
Institute  for Complex Quantum Systems and IQST, Ulm University, Albert-Einstein-Allee 11, 89069  Ulm, Germany}

\author{Joachim Ankerhold}
\email{joachim.ankerhold@uni-ulm.de}
\affiliation{
Institute  for Complex Quantum Systems and IQST, Ulm University, Albert-Einstein-Allee 11, 89069  Ulm, Germany}

\date{\today}

\begin{abstract}
Embedding non-Markovian open quantum dynamics into an enlarged Markovian space offers a powerful route to nonperturbative simulations, where the dynamics of the extended space can be governed by multiple distinct Markovian equations. We show that these distinct embeddings arise from different unravelings of Gaussian bath self-energies, generating a family of deterministic, time-local equations for the extended system. Using the Brownian-oscillator spectral density as an illustrative example, we clarify the relationships among existing approaches, including the Hierarchical Equations of Motion (HEOM) and the Lindblad--pseudomode formalism, and demonstrate how this framework enables numerically stable and efficient simulations. This work provides both a transparent theoretical foundation for embedding techniques and a flexible platform for developing new methods to simulate non-Markovian quantum dynamics.
\end{abstract}

\maketitle

\section{Introduction}
\label{sec:introduction}

Open quantum systems \cite{breuer02,gardiner2004quantum,weiss12} provide a theoretical framework for modeling quantum subsystems interacting with environmental degrees of freedom \cite{breuer16,vega17,xu2026simulating}. They play a central role in a wide range of fields, including quantum optics, condensed-matter physics, chemical physics, quantum reservoir engineering, as well as the development of error mitigation strategies and quantum error-correction protocols \cite{kobayashi2024tensor} in the emerging era of fault-tolerant quantum computing.

Over the past decades, several nonperturbative numerical approaches based on path-integral formulations have been developed. Representative examples include path-integral Monte Carlo \cite{egger2000path,muhlbacher2005nonequilibrium}, the quasi-adiabatic path-integral method \cite{makri1995tensor}, and tensor-network path-integral techniques \cite{strathearn2018efficient,bose2022multisite}. These methods evaluate the influence functional, through which environmental effects are incorporated implicitly via bath correlation functions.

Alternatively, environmental effects can be treated explicitly by representing the bath through discrete modes within time-local equations of motion. For example, approaches such as the multi-configuration time-dependent Hartree method \cite{wang2003multilayer,wang10from,lindoy2021time}, the matrix product state techniques \cite{bulla2005numerical,chin2010exact,ren2022time}, and Davydov ansatz methods \cite{zhao2023the,zeng2025variational,yan2025multiple} employ discretization of environmental modes in the real-frequency domain, often combined with thermofield transformations \cite{yu2004non,borrelli2017simulation,gelin2021efficient,vega2015thermofield} to account for finite-temperature effects. The Lindblad pseudomode formalism \cite{tamascelli2018nonperturbative,pleasance2020generalized,luo2023quantum,lentrodt2023certifying,liang2024purified,park2024quasi,huang2025coupled,alford2025subtleties} introduces damped modes to represent structured reservoirs, while the hierarchical equations of motion (HEOM) \cite{tanimura89,tanimura06,tanimura2020numerically,lambert2019modelling,yan2014theory,ke2022hierarchical,ke2023tree,chen2025tree,li2022low,xu2023universal,takahashi2024finite,su2025nonperturbative,xu2026simulating} incorporate damped modes together with nontrivial system-mode couplings.

Compared with other time-local equation of motion approaches, one distinctive feature of the HEOM is that the initial auxiliary density operators (ADOs) are set to zero, while the reduced density operator is identified with the lowest-tier element, $\hat{\rho}_{0,\ldots,0}$. This structure implies that the auxiliary modes are initialized in their vacuum state, and the reduced density operator is obtained by projecting the total density operator onto the vacuum subspace of these auxiliary modes. Although this property follows directly from the mathematical construction of the ADOs hierarchy \cite{tanimura89}, its physical interpretation has not yet been fully clarified \cite{xu2023universal,xu2026simulating}.

By contrast, in the Lindblad-pseudomode formalism, the initialization of auxiliary modes depends on the decomposition of the bath correlation function. The auxiliary modes may be prepared either in vacuum \cite{tamascelli2018nonperturbative} or thermal states \cite{luo2023quantum}, and the reduced density operator is obtained by performing a partial trace over the auxiliary degrees of freedom. Similarly, matrix product state based approaches combined with thermofield transformation maps finite-temperature environments onto enlarged Hilbert spaces in which the environmental modes are initialized in vacuum states, while thermal fluctuations are encoded in system-mode couplings \cite{vega2015thermofield,xu2026simulating}.

To clarify and explicitly elucidate the connections among these seemingly distinct formulations, \textcite{xu2026simulating} very recently developed a unified  framework termed Quantum Dissipation in Minimally Extended State Space (QD-MESS). This framework serves a dual purpose: By demonstrating how these various non-Markovian approaches emerge from the QD-MESS formulation, it provides a unified picture of non-Markovian approaches with extended state space and thus
provides experts with a systematic tool to select optimal representations based on either physical insight or classical simulation efficiency. Moreover, this conceptual transparency is a prerequisite for integrating corresponding dynamical equations with emerging quantum simulation and computing architectures, leveraging their potential to efficiently tackle non-Markovian open quantum many-body problems.

In this work, we focus on a specific aspect which has not been included in \cite{xu2026simulating}, namely, to further consider the QD-MESS from a path-integral perspective with the goal to provide more explicit simulation schemes. By expressing the Gaussian bath self-energy in different algebraic forms in terms of Green's (response) functions, we demonstrate that the resulting unraveling introduces deterministic auxiliary fields subject to specific boundary conditions. Within this formulation, we show that the HEOM and the Lindblad--pseudomode approach emerge naturally from different choices of Green's function structures. Furthermore, we demonstrate that these seemingly distinct formulations are related through linear transformations in the auxiliary-mode path space, corresponding precisely to Bogoliubov transformations. The associated boundary conditions determine both the preparation of the initial auxiliary states and the procedure for obtaining the reduced density operator.

The remainder of this paper is organized as follows. In Sec.~\ref{Sec:OpenQuantumSystems}, we introduce the open quantum system model, the factorized self-energies, and the Brownian-oscillator spectral density. Section~\ref{sec:qdmess} presents the general framework of Gaussian unraveling, including constructions for different Green's functions and the relations to previous results. Finally, Sec.~\ref{sec:discussion} contains a discussion of the results.

\section{Open Quantum System Modeling}
\label{Sec:OpenQuantumSystems}

We consider an open quantum system coupled linearly to an external environment modeled as a complex-valued Gaussian noise process. The noise effects are completely characterized by its two-point correlation function. The Hamiltonian describing the noise driven system takes the form
\begin{equation}
\hat{H}(t) = \hat{H}_s + \hat{S} \hat{X}_b(t) \,,
\end{equation}
where $\hat H_s$ is the bare system Hamiltonian, $\hat{S}$ is the system coupling operator, and $\hat{X}_b(t)$ denotes time-dependent operator-valued Gaussian noise. Its two-point correlation function (with $\hbar = k_B = 1$) is
\begin{align}
    C(t) = \left\langle \hat{X}_b(t) \hat{X}_b^\dagger(0) \right\rangle_b.
\end{align}
More complicated couplings can be represented as sums of separable terms \cite{vega17}. Importantly, the operator $\hat X_b(t)$ is not assumed to arise from a specific bath model in Hilbert or Liouville space; instead, it is characterized solely by its statistical properties in a stationary state. For concreteness, the Gaussian environment can be realized in Hilbert space as an infinite collection of harmonic oscillators \cite{leggett87,ford1988quantum}.

The Gaussian noise is equivalently characterized by its power spectrum,
\begin{align}
    S_\beta(\omega) = \int_{-\infty}^{\infty} dt\, e^{i\omega t}\, C(t) \,.
\end{align}
The fluctuation-dissipation theorem \cite{Kubo1966the,weiss12} relates the positive and negative frequency components of $S_\beta(\omega)$, which can be expressed as
\begin{equation}\label{Eq:spectralnoise}
S_\beta(\omega) = \frac{J(\omega)}{1- e^{-\beta\omega}} \,.
\end{equation}
Here, $\beta = 1/T$ is the inverse temperature, and $J(\omega)$ is the antisymmetric spectral density, characterized by a finite bandwidth with cutoff frequency $\omega_c$.

We assume that, for a given realization of the stochastic noise
$X_b$, the system density operator evolves according to the
stochastic Liouville equation
\begin{align}
\dot{\hat{\rho}}(t,X_b)
= -i[\hat{H}(t),\hat{\rho}(t,X_b)]
\equiv -i\,\mathcal{L}(t)\,\hat{\rho}(t,X_b),
\end{align}
where $\mathcal{L}(t)$ denotes the Liouvillian associated with the
noise-dependent Hamiltonian.

For an initially uncorrelated system–noise state, the physical reduced
density operator is obtained by averaging over all noise realizations,
\begin{align}
\hat{\rho}_s(t)
&\equiv \bigl\langle \hat{\rho}(t,X_b) \bigr\rangle_b
\notag\\
&=
\left\langle
\mathcal{T}
\exp[-i\!\int_0^t d\tau\,\mathcal{L}(\tau)]
\right\rangle_{\!b}\,
\hat{\rho}_s(0)
\notag\\[0.3em]
&\equiv \mathcal{J}(t)\,\hat{\rho}_s(0),
\end{align}
which defines the noise-averaged propagator $\mathcal{J}(t)$. For an unbiased interaction and Gaussian noise, the propagator in the interaction picture can be evaluated exactly
\cite{kubo62,feynman63,aurel20,huang2024unified},
\begin{align}\label{Eq:sysproggauss}
\mathcal{J}(t)
&=
\mathcal{T}\exp\!\left[
-\frac{1}{2}
\int_0^t d\tau\,du\,
\bigl\langle
\mathcal{T}\mathcal{L}(\tau)\mathcal{L}(u)
\bigr\rangle_b
\right]
\notag\\[0.3em]
&=
\mathcal{T}\,e^{-i\Phi(t)},
\end{align}
where the time-ordered exponential represents the operator form of the
Feynman–Vernon influence functional \cite{feynman63}. The influence phase $\Phi(t)$ can be expressed as
\begin{equation}\label{Eq:PhiSuperop}
\Phi(t)
=
\int_0^t d\tau\,du\,
\begin{bmatrix}
\mathcal{S}_{q}(\tau) & \mathcal{S}_{c}(\tau)
\end{bmatrix}
\bm{\Sigma}(\tau-u)
\begin{bmatrix}
\mathcal{S}_{q}(u) \\[0.2em] \mathcal{S}_{c}(u)
\end{bmatrix},
\end{equation}
where $\bm{\Sigma}$ is the bath self-energy kernel, $\mathcal{S}_{c}$ and $\mathcal{S}_{q}$ denote the classical and quantum superoperators, defined by their action on a density operator as
\begin{equation}
\mathcal{S}_{c/q}\,\hat{\rho}
=
\frac{1}{\sqrt{2}}
\bigl(\hat{S}\hat{\rho}\pm\hat{\rho}\hat{S}\bigr).
\end{equation}

The self-energy matrix $\bm{\Sigma}(t)$ in Eq.~\eqref{Eq:PhiSuperop} is not unique. Exploiting the Hermiticity of the correlation function, $[C(t)]^\dagger = [C(-t)]^\ast = C(t)$, together with the symmetry of the integration domain, one may choose different but equivalent representations \cite{xu2026simulating}. Moreover, $\Phi$ does not depend on the antisymmetric part of the matrix $\Sigma$. For instance, a convenient choice is
\begin{equation}\label{Eq:GeneralSelfEnergy}
    \bm{\Sigma}(t) = 
    \begin{bmatrix}
        -i & -i\,\theta(t) \\
        i\,\theta(-t) & 0
    \end{bmatrix}
    C(t),
\end{equation}
which collects the contributions of the two-point correlation function $C(t)$. Accordingly, Eq.~\eqref{Eq:PhiSuperop} incorporates reservoir dynamics only implicitly, through a retarded self-interaction. The Heaviside step function $\theta(t)$ enforces causality, equaling 1 for $t\ge0$ and 0 otherwise.

An alternative representation of the influence phase recasts the self-energy in a form that explicitly separates real and imaginary parts:
\begin{align} \label{Eq:SelfEnergyRealRI}
\bm{\Sigma}'(t) &= -i \theta(t) \left(
C(t)
\begin{bmatrix}
    1 & 1\\ 0& 0
\end{bmatrix}
+ C^\ast(t)
\begin{bmatrix}
    1 & -1\\ 0& 0
\end{bmatrix}
\right) \notag \\[1ex]
&= -2i \theta(t) \left(
\Re C(t)
\begin{bmatrix}
    1 & 0\\ 0& 0
\end{bmatrix}
+ \Im C(t)
\begin{bmatrix}
    0 & i\\ 0& 0
\end{bmatrix}
\right).
\end{align}
This representation highlights the distinct roles of the real and imaginary parts of the correlation function in generating dissipation and fluctuation effects. The freedom in choosing $\bm{\Sigma}(t)$ reflects the underlying gauge freedom in the embedding construction and will be exploited throughout this work to generate different Markovian embedding schemes.

\subsection{Self-Energy as Matrix Product}
\label{sec:factorization}
The self-energy matrix $\bm{\Sigma}(t)$ introduced in \eref{Eq:PhiSuperop} can be written in several mathematically equivalent forms. Here we make the ansatz that it admits a left–right factorization
\begin{align}\label{Eq:structureUGV}
    \bm{\Sigma}(t) = \mathbb{U}^\dagger\, \mathbb{G}(t)\, \mathbb{V} \,,
\end{align}
where the invertible block matrix $\mathbb{G}(t)$ satisfies an inhomogeneous differential equation
\begin{equation}\label{Eq:inhomoGreen}
    \mathbb{K}(i\partial_\tau)\, \mathbb{G}(\tau) = \delta(\tau)\,\mathds{1} 
\end{equation}
where $\mathbb{K}$ is the sum of a  a constant matrix and a matrix whose non-zero entries are $\pm i\partial_t$. This ansatz can be made exact for a multi-exponential correlation function $C(t)$. Otherwise, it approximates $C(t)$ to any given accuracy when the rank $K$ of the matrix $\mathbb{G}$ is raised sufficiently.
The time-independent matrices $\mathbb{U}^\dagger$ and $\mathbb{V}$ are introduced without imposing any additional structure or constraints.

The factorization in \eref{Eq:structureUGV} is evidently not unique, even for a fixed rank $K$ of $\mathbb{G}(t)$. Any pair of invertible linear transformations $\mathbb{L}$ and $\mathbb{R}$ acting on $\mathbb{U}^\dagger$ and $\mathbb{V}$ generates an equivalent triplet
\begin{equation}\label{Eq:LinearTransformation}
    \mathbb{U}'^\dagger = \mathbb{U}^\dagger\,\mathbb{L}^{-1}, \quad
    \mathbb{G}' = \mathbb{L}\, \mathbb{G}\, \mathbb{R}^{-1}, \quad
    \mathbb{V}' = \mathbb{R}\, \mathbb{V},
\end{equation}
which leaves the product in \eref{Eq:structureUGV} invariant. This gauge-like freedom means that numerous mathematically different choices of $(\mathbb{U},\mathbb{G},\mathbb{V})$ correspond to the same self-energy, although they may lead to distinct physical interpretations or computational schemes (in face of single-particle basis truncation).

\subsection{Brownian-Oscillator Spectral Density}

By way of example, we will illustrate the general formalism by considering a class of bath correlation functions of the form
\begin{align}\label{Eq:CorrelationDecomposition}
    C(t \ge 0) =
    c_1\, e^{-i(\zeta - i\gamma_0) t}
    + c_2\, e^{i(\zeta + i\gamma_0) t} \,,
\end{align}
where $c_{1,2}$ are complex coefficients and $\zeta,\gamma_0$ are real parameters.
This structure describes a damped mode with complex frequencies
$\pm\zeta - i\gamma_0$ located in the lower half of the complex plane.
The self-energy can be written as
\begin{align}\label{Eq:SelfEnergyMatrix}
\Sigma(t)
&=\!
\begin{bmatrix}
(c_1 \!+\! c_2^\ast) G^R(t) - (c_1^\ast \!+\! c_2) G^A(t)
&
(c_1 \!-\! c_2^\ast) G^R(t)
\\[6pt]
(c_1^\ast \!-\! c_2) G^A(t)
& 0
\end{bmatrix}
\end{align}
with
\begin{subequations}\label{Eq:Gretadv}
\begin{align}
G^R(t) &= -i\,\theta(t)\,e^{-i(\zeta-i\gamma_0)t} \,, \\
G^A(t) &= \;\, i\,\theta(-t)\,e^{\,i(\zeta+i\gamma_0)t} \,.
\end{align}    
\end{subequations}
Eq.~\eqref{Eq:SelfEnergyMatrix} emerges after inserting $C(t)$ into Eq.~\eqref{Eq:GeneralSelfEnergy}, with some modifications allowed by symmetries. One of the many possible decompositions of type~\eqref{Eq:structureUGV} is
\begin{equation}
\begin{aligned}
\mathbb{U}^\dagger &=
\begin{bmatrix}
1 & c_2 - c_1^\ast \\
0 & c_1^\ast - c_2
\end{bmatrix},\quad
\mathbb{V} =
\begin{bmatrix}
1 & 0 \\
c_1 + c_2^\ast & c_1^\ast - c_2
\end{bmatrix},\\
\mathbb{G}(t) &=
\begin{bmatrix}
0 & G^R(t)\\
G^A(t) & 0
\end{bmatrix}.
\end{aligned} \notag
\end{equation}
A more general form of this matrix product is related to the Keldysh formalism discussed in  Sec.~\ref{sec:qdmess}.
The generalization of Eq.~\eqref{Eq:CorrelationDecomposition} to a sum of multiple exponential terms is straightforward.

A minimal physical model that realizes the decomposition in Eq.~\eqref{Eq:CorrelationDecomposition} is the widely studied Lorentzian-type spectral density \cite{garg1985effect,grabert1984quantum,hanggi2005fundamental,meier1999non,tanaka2009quantum,li2022low,risken84,weiss12}
\begin{align}\label{Eq:brownianspectral}
    J(\omega) = \frac{4 c_0^2 \gamma_0 \omega}
    {(\omega^2 - \omega_0^2)^2 + 4 \gamma_0^2 \omega^2}\, .
\end{align}
It typically arises as an effective spectral density when a broadband Ohmic bath is filtered by a single harmonic mode at frequency $\omega_0$ \cite{garg1985effect}. The corresponding correlator, across different parameter regimes, captures a broad range of interesting physical processes. In the sequel, we will use at several places three limiting cases to specify general results. 

\emph{Case~1: Classical contribution.--}
At sufficiently elevanted temperatures, when Matsubara terms can be neglected, one regains the classical part of the correlation function 
\begin{align}\label{Eq:ClassicalBrownianCorrelationFunction}
    C(t) &= \frac{c_0^2}{4\zeta}
    \left[
    \coth\!\left(\frac{\beta(\zeta - i\gamma_0)}{2}\right) + 1
    \right] e^{-i(\zeta-i\gamma_0)t} \notag \\
    &\quad + \frac{c_0^2}{4\zeta}
    \left[
    \coth\!\left(\frac{\beta(\zeta + i\gamma_0)}{2}\right) - 1
    \right] e^{-i(\zeta+i\gamma_0)t} \,,
\end{align}
where $\zeta = \sqrt{\omega_0^2 - \gamma_0^2}$ is the effective oscillation frequency.

\emph{Case~2: Weak-damping regime.--}
In the weak-damping limit $\gamma_0 \ll \omega_0,\zeta$, where the spectral distribution is sharply peaked, the correlation function \eqref{Eq:ClassicalBrownianCorrelationFunction} reduces to the quasi-thermal form 
\begin{align}\label{Eq:quasiThermalDecomposition}
    C(t \ge 0)
    =
    \frac{c_0^2}{2\zeta}
    \left[
    (n_\beta + 1)e^{-i\zeta t}
    + n_\beta e^{i\zeta t}
    \right]
    e^{-\gamma_0 t} \,,
\end{align}
where $n_\beta = (e^{\beta\zeta} - 1)^{-1}$ is the Bose–Einstein occupation number. This spectral density is widely used in quantum optics and also serves as an effective description of a discrete bath mode with a narrow linewidth broadening $\gamma_0$. Moreover, it serves as a basis function for fitting bath correlation functions in the Lindblad–pseudomode formulation \cite{somoza2019dissipation,luo2023quantum,leppakangas2023quantum} .

\emph{Case~3: Strong-friction limit.--}
In the overdamped regime $\gamma_0 \gg \omega_0$, the correlation function \eqref{Eq:ClassicalBrownianCorrelationFunction} for times
$t \gtrsim \omega_D^{-1} \gg (2\gamma_0)^{-1}$ reduces to
\begin{align}\label{Eq:DebyeCorrelationFunction}
    C(t) \approx
    \frac{c_0^2}{4\gamma_0}
    \left(\cot\!\frac{\beta\omega_D}{2}-i\right)
    e^{-\omega_D t} \,,
\end{align}
where $\omega_D = \omega_0^2/(2\gamma_0)$ is the Drude cutoff frequency.
This form corresponds to the Debye spectral density, widely used in HEOM studies in chemical physics \cite{tanimura89,tanimura2020numerically,bai2024hierarchical}.

\section{Gaussian Unraveling of the Influence Functional}
\label{sec:qdmess}

Now, we proceed with the general discussion of Sec.~\ref{Sec:OpenQuantumSystems}.
Substituting \eref{Eq:structureUGV} into \eref{Eq:PhiSuperop}, the influence phase takes the compact bilinear form
\begin{align}\label{Eq:phiDecomposition}
    \Phi(t) &=\! \int_0^t\!\! dt_1 dt_2
    \begin{bmatrix}
        \mathbb{\mathcal{S}}_q(t_1) \!&\! \mathbb{\mathcal{S}}_c(t_1)
    \end{bmatrix} 
    \mathbb{U}^\dagger 
    \mathbb{G}(t_1 \!-\! t_2)\,
    \mathbb{V} 
    \begin{bmatrix}
        \mathbb{\mathcal{S}}_q(t_2) \\[2pt] \mathbb{\mathcal{S}}_c(t_2)
    \end{bmatrix} \notag \\ 
    &\equiv \! \int_0^t\!\! dt_1 dt_2\,  \mathcal{S}_u^\dagger(t_1)\, \mathbb{G}(t_1 \!-\! t_2)\, \mathcal{S}_v(t_2) \,.
\end{align}
Here the effect of the matrices $\mathbb{U}^\dagger$ and $\mathbb{V}$ has been absorbed into the source fields by defining the row and column superoperator vectors: 
\begin{align}
    \mathcal{S}_u^\dagger = \begin{bmatrix}
        \mathcal{S}_q^\dagger & \mathcal{S}_c^\dagger
    \end{bmatrix} \mathbb{U}^\dagger,\quad 
    \mathcal{S}_v = \mathbb{V} \begin{bmatrix}
        \mathcal{S}_q & \mathcal{S}_c
    \end{bmatrix}^T.
\end{align}
We emphasize that the operator-valued influence phase acting on the system is equivalent to that in \eref{Eq:PhiSuperop}; it has merely been recast into different structures for computing the inversion of Green's function and obtaining the boundary constraints for the unraveling fields. 

Our objective is to recast the influence functional, whose phase is intrinsically time-nonlocal, into a time-local representation. This allows us to derive deterministic, time-local equations of motion while keeping explicit track of the boundary conditions. Such a transformation can be achieved by using the Gaussian identity \cite{altland2010condensed,sieberer2016keldysh}
\begin{equation}
\resizebox{\columnwidth}{!}{$
\begin{aligned}
&\int\! \mathcal{D}[{\Psi}^\dagger,{\Psi}] 
\exp{
i\!\int_0^t\!\! d\tau 
\Bigl[
{\Psi}^\dagger(\tau) \mathbb{K}(i\partial_\tau) {\Psi}(\tau)
- {\Psi}^\dagger(\tau) \mathcal{S}_{v}(\tau)
- \mathcal{S}_{u}^\dagger(\tau) {\Psi}(\tau)
\Bigr]} \\
&= \exp{
-i\!\int_0^t\!\! d\tau du\,
\mathcal{S}_{u}^\dagger(\tau)\, \mathbb{G}(\tau-u)\, \mathcal{S}_{v}(u) } \,. \notag
\end{aligned}
$}
\end{equation}
The auxiliary field $\Psi(t)$ is a column vector whose dimension matches that of $\mathbb{K}(i\partial_t)$, and $\Psi^\dagger(t)$ denotes its Hermitian conjugate. The path-integral normalization factor has been absorbed into the functional measure $\mathcal{D}[{\Psi}^\dagger,{\Psi}]$. Implementing the Gaussian identity to \eqref{Eq:sysproggauss} with respect to influence phase representation \eqref{Eq:phiDecomposition}, the reduced density operator is then expressed as a path integral over auxiliary fields:
\begin{align}\label{Eq:generalPathIntegral}
    \hat\rho_s(t) = \mathcal{T}\! \int\mathcal{D}[\Psi^\dagger,\Psi]\, e^{i\mathcal{A}[\mathcal{S}_u^\dagger,\mathcal{S}_v;\Psi^\dagger,\Psi]}\, \hat\rho_s(0)  \,,
\end{align}
with the time-local action 
\begin{align}\label{Eq:timeLocalActionGeneral}
\mathcal{A}[\mathcal{S}_u^\dagger,\mathcal{S}_v;\Psi^\dagger,\Psi] &= \!\int_0^t\! d\tau \Big[ {\Psi}^\dagger(\tau)\, \mathbb{K}(i\partial_\tau)\, {\Psi}(\tau)  \notag \\
& \quad - {\Psi}^\dagger(\tau)\, \mathbb{\mathcal{S}}_v(\tau) - \mathbb{\mathcal{S}}_u^\dagger(\tau)\, {\Psi}(\tau)  \Big] \,.
\end{align}
The Schr\"odinger equivalent of this double path action does not translate into unitary time evolution; some terms may be identified as Lindblad dissipators or other non-Hamiltonian terms.
Construction of the proper initial states and terminal weights for paths can be accomplished by carefully examining the continuum limit of time-discrete path integrals. The validity of the corresponding initial preparation and terminal projection operations can be verified by tracing out the auxiliary fields, recovering the superoperator form of the influence functional~\cite{aurel20}.

In the interaction picture for system plus auxiliary modes, the total density matrix evolves along a Gaussian trajectory according to
\begin{align}\label{Eq:evolutionTrajectory}
\dot{\hat{\rho}}(t)
&= -i\mathcal{L}_{\rm sm}(t) \hat{\rho}(t) \notag \\
&= -i\left[ \Psi^\dagger(\tau)\,\mathbb{\mathcal{S}}_v(\tau) + \mathbb{\mathcal{S}}_u^\dagger(\tau)\,\Psi(\tau)
\right]\hat{\rho}(t) \,.
\end{align}

In the following, we focus for the correlator in Eq.~(\ref{Eq:CorrelationDecomposition}) on two prototypical structures of the Green’s function $\mathbb{G}$, namely the Keldysh and the decoupled block forms,  which give rise to distinct deterministic, time-local equations of motion; the term involving $\mathbb{K}(i\partial_\tau)$ plays the role of the Lagrangian for a set of free harmonic modes, whose dynamics are governed by the corresponding homogeneous equations of motion $\mathbb{K}(i\partial_t)\, {\Psi}(t) = \mathbf{0}$. The boundary conditions of the fields follow directly from the discrete real-time representation of $\mathbb{K}(i\partial_t)$ as computed from $\mathbb{G}(t)$. 

\subsection{Keldysh Path-Integral Representation}
 We factorize the self-energy~\eqref{Eq:SelfEnergyMatrix} in the form of Eq.~\eqref{Eq:structureUGV} using the $2\times2$ Keldysh Green's function matrix
\begin{equation}\label{Eq:kmatrix}
\mathbb{G}(t)=
  \begin{bmatrix}
        G^{K}(t) & G^{R}(t) \\
        G^{A}(t) & 0
  \end{bmatrix}.
\end{equation}
The retarded and advanced components~\eref{Eq:Gretadv} describe causal propagation of a
damped bosonic mode with frequency $\zeta$ and damping $\gamma_0$,
while the Keldysh component encodes the fluctuation level through the reference occupation $n_{\rm ref}$, 
\begin{equation}
G^K(t)=(2n_{\rm ref}+1)\bigl[G^R(t)-G^A(t)\bigr] \,.
\end{equation}
Here $n_{\rm ref}$ specifies the Gaussian reference occupation of the
auxiliary mode; in particular, $n_{\rm ref}=0$ corresponds to a vacuum
reference state.

The $2\times 2$ matrices $\mathbb{U}^\dagger$ and $\mathbb{V}$ appearing in the factorization \eqref{Eq:structureUGV} encode the remaining freedom in the representation: 
\begin{subequations}
\begin{align}
\mathbb{U}_{\rm KP}^\dagger
&=\!
\begin{bmatrix}
\delta & \delta(2n_{\rm ref}+1)-(c_1^\ast+c_2)/\lambda \\
0 & (c_1^\ast-c_2)/\lambda 
\end{bmatrix}, \\[1ex]
\mathbb{V}_{\rm KP}
&=\!
\begin{bmatrix}
\lambda & 0 \\
(c_1+c_2^\ast)/\delta - \lambda(2n_{\rm ref}+1)
&
(c_1-c_2^\ast)/\delta
\end{bmatrix},
\end{align}
\end{subequations}
with nonzero free parameters $\delta$ and $\lambda$. When $c_1^\ast + c_2 \in \mathbf{R}$, the off-diagonal terms can be removed by imposing the constraint $(2n_{\rm ref}+1)\delta\lambda = c_1^\ast + c_2$, which yields a simplified form of the generator $\mathcal{L}_{\rm sm}$. 

In the steady state, time-translation invariance of the Green's function implies that its inverse kernel can be written as a local differential operator (up to boundary contributions at the contour endpoints), 
\begin{equation}\label{Eq:inverseGreenMatrixKeldysh}
\mathbb{K}(i\partial_t)=
\begin{bmatrix}
0 & i\partial_t-(\zeta+i\gamma_0) \\
i\partial_t-(\zeta-i\gamma_0) & 2i\gamma_0(2n_{\rm ref}+1)
\end{bmatrix}.
\end{equation}

The structure of Keldysh Green's function \eqref{Eq:kmatrix} specifies the unraveled path integral representation \eqref{Eq:generalPathIntegral} in terms of complex-valued Keldysh fields $\Psi^\dagger(t)=[\phi_c^\dagger(t)\ \phi_q^\dagger(t)]$ and $\Psi(t)=[\phi_c(t)\ \phi_q(t)]^T$. These fields are obtained from the forward and backward contour fields $\phi_\pm(t)$ via the Keldysh rotation~\cite{kamenev2023field,sieberer2016keldysh}
\begin{equation}
\phi_{c/q}(t)=\frac{1}{\sqrt{2}}\bigl[\phi_{+}(t)\pm\phi_{-}(t)\bigr]\,,
\end{equation}
where $\phi_\pm$ correspond to coherent-state eigenvalues of bosonic annihilation superoperators acting on the density matrix, $\hat a_{\pm}\ket{\phi_\pm}=\phi_\pm\ket{\phi_\pm}$ in vectorized form, with $\hat a_{+}$ and $\hat a_{-}$ acting from the left and right, respectively on the density matrix $\rho(t)$~\cite{xu2026simulating}. The classical and quantum superoperators are defined as $\hat{a}_{c/q} =  (\hat{a}_+ \pm \hat{a}_-)/\sqrt{2}$. 

Substituting the kernel~\eqref{Eq:inverseGreenMatrixKeldysh} into Eq.~\eqref{Eq:timeLocalActionGeneral} yields an effective generator $\mathcal{L}_{\rm sm}$ that governs the coupled dynamics of the system and the auxiliary modes. By carefully accounting for the boundary conditions of $\mathbb{G}(t)$ and its inverse, the path integral  representation \eqref{Eq:generalPathIntegral} can be transformed into an equivalent operator formulation, in which the reduced density matrix is obtained by tracing out the auxiliary mode. In the interaction picture,
\begin{align}\label{Eq:PI2EvolutionTrace}
\hat{\rho}_s(t)
&= \Tr_{\rm b}\{\hat{\rho}(t)\} \notag\\
&= \Tr_{\rm b}\!\left\{
\mathcal{T}\exp\!\left[-i\!\int_0^t d\tau\,\mathcal{L}_{\rm sm}(\tau)\right]
\hat{\rho}(0)
\right\} \,,
\end{align}
assuming a factorized initial state
$\hat{\rho}(0)=\hat{\rho}_s(0)\otimes\hat{\rho}_m(0)$,
where the auxiliary mode is prepared in a thermal state with occupation
$n_{\rm ref}$.

Replacing the auxiliary fields by creation and annihilation operators in
Eq.~\eqref{Eq:evolutionTrajectory} yields the Schr\"odinger picture
equation of motion (setting $\delta=\lambda=1$),
\begin{align}\label{Eq:GeneralLindbladpseudomodeEquation}
\dot{\hat{\rho}}(t)
&= -i\mathcal{L}_{\rm sm}\hat{\rho}(t) \notag\\
&= -i\mathcal{L}_0\hat{\rho}(t)
+ \gamma_0(n_{\rm ref}+1)
\Bigl(2\hat a\,\hat\rho\,\hat a^\dagger
- \{\hat a^\dagger\hat a,\hat\rho\}\Bigr)
\notag\\
&\quad
+ \gamma_0 n_{\rm ref}
\Bigl(2\hat a^\dagger\hat\rho\,\hat a
- \{\hat a\hat a^\dagger,\hat\rho\}\Bigr),
\end{align}
which reveals that the original system–bath model with correlation function~\eqref{Eq:CorrelationDecomposition} admits an equivalent representation as Markovian dynamics of an enlarged system governed by a Lindblad generator. The bare system-auxiliary-modes dynamics is governed by the Liouvillian 
\begin{align}\label{Eq:GeneralLiouvilleLindbladPseudomode}
\mathcal{L}_0
&= [\hat H_s + \zeta\hat{a}^\dagger\hat{a},\cdot]
+
\begin{bmatrix}
\mathcal{S}_q^\dagger & \mathcal{S}_c^\dagger
\end{bmatrix}
\mathbb{U}_K^\dagger
\begin{bmatrix}
\hat a_c \\ \hat a_q
\end{bmatrix}  \notag \\
&\quad +
\begin{bmatrix}
\hat a_c^\dagger & \hat a_q^\dagger
\end{bmatrix}
\mathbb{V}_K
\begin{bmatrix}
\mathcal{S}_q \\ \mathcal{S}_c
\end{bmatrix}.
\end{align}

We emphasize that the partial trace operation as Eq.~\eqref{Eq:PI2EvolutionTrace} together with the initial state $\hat{\rho}_m(0)$, arise from the structure of Green’s function in Eq.~\eqref{Eq:kmatrix}. The auxiliary-mode initial state therefore plays the role of a Gaussian reference state that fixes the fluctuation level of the representation. It is not tied to the physical temperature of the target environment, whose influence enters separately through the correlation decomposition in Eq.~\eqref{Eq:CorrelationDecomposition}.

\emph{Example~1.—}
In the parameter regime corresponding to Eq.~\eqref{Eq:quasiThermalDecomposition} and choosing $n_{\rm ref}=n_\beta$, the embedded system Liouvillian reduces to
\begin{align}
    \mathcal{L}_0 =
    \bigl[\hat{H}_s
    + \frac{c_0}{\sqrt{2\zeta}}\hat{S}(\hat{a}+\hat{a}^\dagger)
    + \zeta\hat{a}^\dagger\hat{a},\,\cdot\bigr] \,.
\end{align}
In this limit, Eq.~\eqref{Eq:GeneralLindbladpseudomodeEquation} recovers  the conventional Lindblad-pseudomode equation~\cite{luo2023quantum,tamascelli2018nonperturbative,pleasance2020generalized} with Hamiltonian system-mode coupling. The auxiliary mode is coherently coupled to the system and undergoes Markovian damping described by the Lindblad dissipator, thereby reproducing the original bath correlation function. This picture is close to the modeling of open quantum systems through a shift in the system-environment divide~\cite{garg1985effect}.

\emph{Example~2.--} For the correlators \eqref{Eq:ClassicalBrownianCorrelationFunction} and \eqref{Eq:quasiThermalDecomposition} within the limit of $\gamma_0 = 0$, we select the parameters 
\begin{equation}
    n_{\rm ref} = 0,\quad \delta = \sqrt{\frac{c_0^2}{2\omega}\coth{\frac{\beta\omega}{2}}},\quad \lambda = \frac{c_1^\ast + c_2}{\delta} \,.
    \end{equation}
This choice simplifies the Liouvillian \eqref{Eq:GeneralLiouvilleLindbladPseudomode} to 
\begin{align}\label{Eq:StableHeomDiscrete}
    \mathcal{L}_0 =&\, [\hat{H}_s +\omega_0\hat{a}^\dagger\hat{a},\cdot] + \sqrt{\frac{c_0^2}{2\omega_0}\coth{\frac{\beta\omega_0}{2}}} \hat{S}_q(\hat{a}_c +\hat{a}_c^\dagger)  \notag \\
    &+ \sqrt{\frac{c_0^2}{2\omega_0}\tanh{\frac{\beta\omega_0}{2}}} \hat{S}_c(\hat{a}_q +\hat{a}_q^\dagger) \,.
\end{align}
We define the ADOs as $\hat{\rho}_{k,l} = (l|\hat{\rho}(t)|k)$ in the unnormalized auxiliary modes Fock basis, $(l|\hat{a}^\dagger = il\,(l-1|$, $\hat{a}|k) = -k\,|k-1)$, the resulting dynamical equations for the ADOs become identical to Eq.~(22) of Ref.~\cite{yan2020new}. The boundary conditions for the auxiliary modes are naturally enforced by the ADO dynamics: choosing $n_{\rm ref} = 0$ initializes all ADOs ($m,n \neq 0$) to zero, and the partial trace over the auxiliary modes provides the method to extract the reduced system density operator.

As reported in Refs.~\cite{dunn2019removing,krug23}, the HEOM for discrete modes can suffer from numerical instability under truncation at long evolution times. A similar issue may arises in the HEOM formulations discussed in subsequent two sections due to basis truncation. However, \textcite{yan2020new} demonstrated that transforming these conventional structured HEOM into a representation such as \eqref{Eq:GeneralLindbladpseudomodeEquation} can effectively eliminate this numerical instability.

\subsection{Pure-State Path-Integral Representation}
\label{subsec:frame2}

An alternative and conceptually distinct representation emerges when the self-energy is expressed in a basis that renders the $2\times2$ Green's function matrix $\mathbb{G}(t)$ block diagonal:
\begin{equation}\label{Eq:diagonalGreenMatrix}
\mathbb{G}(t)=
\begin{bmatrix}
G(t) & 0 \\
0 & -G^\dagger(-t)
\end{bmatrix}.
\end{equation}
Here, $G(t) = -i\,\theta(t)\,e^{-i(\zeta-i\gamma_0)t}$ describes retarded propagation of a damped bosonic mode, while $G^\dagger(-t)$ can be interpreted as the corresponding advanced propagator. Equivalently, this structure can be viewed as representing the time-ordered and anti-time-ordered propagators of a damped bosonic mode at zero temperature. The inverse kernel (up to boundary contributions at the contour endpoints) reads 
\begin{align}
\mathbb{K}(i\partial_t) =
\begin{bmatrix}
i\partial_t - (\zeta-i\gamma_0) & {0} \\
{0} &
- i\partial_t + (\zeta + i\gamma_0)
\end{bmatrix}.
\end{align}

The block-diagonal structure of $\mathbb{G}(t)$ decouples the forward and backward sectors, allowing them to be unraveled independently. Their mutual quantum correlations are no longer encoded in their free propagator ${G}(t)$ but are instead mediated entirely through their couplings to the system, as defined by $2\times 2$ matrices $\mathbb{U}^\dagger$ and $\mathbb{V}$:
\begin{equation}
\mathbb{U}_{\rm PS}^\dagger =
\begin{bmatrix}
\delta & \dfrac{c_1^\ast+c_2}{\lambda} \\[3mm]
0 & \dfrac{c_2-c_1^\ast}{\lambda}
\end{bmatrix},\,
\mathbb{V}_{\rm PS} =
\begin{bmatrix}
\dfrac{c_1 + c_2^\ast}{\delta}
&
\dfrac{c_1 - c_2^\ast}{\delta} \\[3mm]
\lambda & 0
\end{bmatrix}. 
\end{equation}

The block-diagonal structure of the Green's function matrix \eqref{Eq:diagonalGreenMatrix} naturally accommodates unraveling fields defined on the forward and backward contours, $\Psi^\dagger(t) = [{\phi}_+^\dagger(t)\; {\phi}_-^\dagger(t)]$ and $\Psi(t) = [{\phi}_+(t)\; {\phi}_-(t)]^T$. Substituting the corresponding kernel $\mathbb{K}(i\partial_t)$ into Eq.~\eqref{Eq:timeLocalActionGeneral} yields an effective generator $\mathcal{L}_{\rm sm}$ governing the coupled system-auxiliary dynamics. By carefully accounting for the boundary conditions of $\mathbb{G}(t)$ and its inverse, the path integral representation of Eq.~\eqref{Eq:generalPathIntegral} can be transformed into an equivalent operator formulation, where the reduced density operator is obtained via a {\em vacuum projection} rather than a partial trace:
\begin{align}\label{PI2EvolutionVacuum}
\hat{\rho}_s(t)
&= \langle {0} | \hat{\rho}_{\rm sm}(t) | {0} \rangle \notag \\
&= \Big\langle {0} \Big| \mathcal{T}
\exp\!\left[-i\!\int_0^t d\tau\, \mathcal{L}_{\rm sm}(\tau)\right]
\hat{\rho}(0)
\Big| {0} \Big\rangle,
\end{align}
assuming a factorized initial state $\hat\rho(0) = \hat\rho_s(0) \otimes \hat{\rho}_{m}(0)$ with the auxiliary mode prepared in its vacuum, $\hat{\rho}_{m}(0) = |{0}\rangle\langle {0}|$. 

Substituting the above obtained triplet ($\mathbb{U}^\dagger, \mathbb{K}(i\partial_t), \mathbb{V}$) into \eqref{Eq:timeLocalActionGeneral}, and replacing the auxiliary fields by creation and annihilation operators in
Eq.~\eqref{Eq:evolutionTrajectory} yields the Schr\"odinger picture
equation of motion, 
\begin{align}\label{Eq:qdmessDiagonal}
\dot{\hat{\rho}}(t) &= -i[\hat{H}_s, \hat\rho(t)] - (\gamma_0 + i\zeta) \hat{a}^\dagger\hat{a} \hat{\rho} - (\gamma_0 - i\zeta) \hat{\rho} \hat{a}^\dagger\hat{a}  \notag \\
& -i \delta\, \mathcal{S}_q \hat{a} \hat{\rho} -i\left(\frac{c_1 + c_2^\ast}{\delta}\, \mathcal{S}_q + \frac{c_1 - c_2^\ast}{\delta}\, \mathcal{S}_c \right) \hat{a}^\dagger\hat{\rho}
\notag\\
&-i \lambda\, \mathcal{S}_q \hat{\rho} \hat{a}^\dagger -i \left(\frac{c_2 + c_1^\ast}{\lambda}\, \mathcal{S}_q + \frac{c_2 - c_1^\ast}{\lambda}\, \mathcal{S}_c \right) \hat{\rho} \hat{a} \,.
\end{align}

\emph{Example~1.--} In the limit $\gamma_0 = 0$, the correlator of Eq.~\eqref{Eq:ClassicalBrownianCorrelationFunction} leads to numerical instability in \eref{Eq:qdmessDiagonal} when the auxiliary-mode basis is truncated \cite{dunn2019removing,yan2020new}. However, this instability can be cured by an appropriate Bogoliubov transformation. Choosing the free parameters $\lambda=\delta=[\frac{c_0^2}{2\omega_0}\coth{\frac{\beta\omega}{2}}]^{\frac{1}{2}}$ for \eref{Eq:qdmessDiagonal} and applying the Liouville space transformation \cite{xu2023universal,xu2026simulating} 
\begin{align}
\mathcal{B} = e^{-\hat{a}_+ \hat{a}_-^\dagger} e^{i\pi \hat{a}_-^\dagger\hat{a}_-} \,,
\end{align}
yields a new time-local equation of motion for $\mathcal{B}\hat{\rho}(t)$ with generator given by Eq.~\eqref{Eq:StableHeomDiscrete}. In this transformed frame, truncation of the auxiliary-mode basis exhibits stable numerical behavior for discrete modes, as demonstrated in Ref.~\cite{yan2020new}.

\emph{Example~2.--}
Choosing the free parameters $\lambda=\delta=\sqrt{2}$, the resulting Liouville equation in the unnormalized Fock basis, $\hat{a}\ket{n}=n\ket{n-1}$, defines the ADOs $\hat{\rho}_{m,n}(t)=(m|\hat{\rho}(t)|n)$. For correlator \eqref{Eq:ClassicalBrownianCorrelationFunction}, equation \eqref{Eq:qdmessDiagonal} reproduces the classical part of
Eq.~(A4) in Ref.~\cite{liu14}.

Alternatively, the ADO array can be reshaped by adopting the convention $\hat{\rho}_{m,n}(t)=(n|\hat{\rho}(t)|m)$, which corresponds to exchanging the auxiliary indices. In the quasi-thermal limit~\eqref{Eq:quasiThermalDecomposition} with $\gamma_0=0$, Eq.~\eqref{Eq:qdmessDiagonal} then reduces to Eq.~(C1) of
Ref.~\cite{liu14}.

\emph{Example~3.--} For the correlator in Eq.~\eqref{Eq:CorrelationDecomposition} with $c_2 = 0$, we choose parameters $\delta = \sqrt{2 c_1}$ and $\lambda = \sqrt{2 c_1^\ast}$. The resulting equation, when represented in the normalized Fock basis defined by $\hat{a}|n\rangle = \sqrt{n}|n-1\rangle$, generates auxiliary density operators $\hat{\rho}_{m,n}(t) = \langle m |\hat{\rho}(t)| n \rangle$. This construction yields the HEOM structure \cite{xu2022taming,suess2015hierarchical,link2022non,vilkoviskiy2024bound,yan2004hierarchical,liang2024purified,zhang2025purified,su2025nonperturbative}.

\emph{Example~4.--} For the symmetric choice $c_1 = c_2 = c_0$ and $\lambda = -\delta$, the correlation function simplifies to 
\begin{align}
    C(t\ge 0) = c_0\, e^{-\gamma t} \cos{\zeta t} \,.
\end{align}
The auxiliary mode in Eq.~\eqref{Eq:qdmessDiagonal} then exhibits left–right symmetry in its coupling to the system Liouville space operators. This symmetry enables a mixed {\em Liouville-Hilbert} space representation: the system evolves in Liouville space, while the auxiliary bosonic mode remains a conventional Hilbert-space operator, coupling to the system from either side, e.g.,
\begin{align}\label{Eq:ReducedqdmessDiagonal}
\dot{\hat{\rho}}(t) &= -i[\hat{H}_s, \hat\rho(t)] - (\gamma_0 + i\zeta) \hat{a}^\dagger\hat{a} \hat{\rho} -i \delta\, \mathcal{S}_q \hat{a} \hat{\rho}  \notag \\
&\quad  -i\left(\frac{c_1 + c_2^\ast}{\delta}\, \mathcal{S}_q + \frac{c_1 - c_2^\ast}{\delta}\, \mathcal{S}_c \right) \hat{a}^\dagger\hat{\rho} \,.
\end{align}
Projecting the equation (setting $\zeta = 0$) onto Fock basis give rise to the conventional HEOM  structure\cite{tanimura89,ishizaki05,shi2009electron} for Debye spectral density with correlator \eqref{Eq:DebyeCorrelationFunction}.

\emph{Example~5.--}
In the limit of strong damping $\gamma_0\gg \omega_0$
\begin{subequations}
    \begin{align}
    \gamma_0 + i\zeta &= \gamma_0 + i\sqrt{\omega_0^2 - \gamma_0^2} \approx \frac{\omega_0^2}{2\gamma_0}  \\
    \gamma_0 - i\zeta &= \gamma_0 - i\sqrt{\omega_0^2 - \gamma_0^2} \approx 2\gamma_0 \,.
\end{align}
\end{subequations}
In time $t \ge (2\gamma_0)^{-1}$, the $\gamma_0-i\zeta$ term becomes the dominant, any terms that creates a right excitation are suppressed by the large $2\gamma_0$ rate, which produces a singular, one-sided projection: the equation collapses onto a right-vacuum manifold. Equation \eqref{Eq:qdmessDiagonal} on timescales $t \ge (2\gamma_0)^{-1}$ collapses to the equation that identity to \eref{Eq:ReducedqdmessDiagonal}, and therefore leads to the conventional HEOM description.

\subsection{Hilbert Space Path-Integral Representation}
\label{sec:hspir}
The unraveling of a Gaussian reservoir based on Eqs.~\eqref{Eq:SelfEnergyRealRI}---\eqref{Eq:inhomoGreen} can also be accomplished using \emph{only} a
 retarded Green’s function, defined as
\begin{align}
    \mathbb{G}(t) = -i\theta(t)e^{-i\bm{\mathcal{E}} t} \,.
\end{align}

In this representation, assuming Gaussian bath correlation function satisfies \cite{ikeda2020generalization}   
\begin{subequations}\label{Eq:RICorrelatorDecomposition}
    \begin{align}
    \Re{C(t\ge 0)} &= i\bm{k}^\dagger \mathbb{G}(t) \bm{\eta} \\
    \Im{C(t\ge 0)} &= i\bm{k}^\dagger \mathbb{G}(t) \bm{\eta}'
\end{align}
\end{subequations}
with $K$ dimensional column vectors $\bm{\kappa}$, $\bm{\eta}$, and $\bm{\eta}'$. The self-energy \eqref{Eq:SelfEnergyRealRI} admits a left–right factorization
\begin{align}\label{Eq:HilbertSelfEnergy}
\Sigma(t)
&=
\begin{bmatrix}
\sqrt{2}\bm{\kappa}^\dagger\\
\bm{0}
\end{bmatrix}
\mathbb{G}(t)
\begin{bmatrix}
\sqrt{2}\bm{\eta} & i\sqrt{2}\bm{\eta}'
\end{bmatrix}  \notag \\
& \equiv
\mathbb{U}_{\rm HS}^\dagger\, \mathbb{G}(t) \, \mathbb{V}_{\rm HS} \,.
\end{align}

The inverse Green's function corresponding to $G(t)$ is given by
\begin{align}
\mathbb{K}(i\partial_t)=i\partial_t-\bm{\mathcal{E}} \,,
\end{align}
which determines the path-integral representation in Eq.~\eqref{Eq:generalPathIntegral} and the corresponding generator $\mathcal{L}_{\rm sm}$ in Eq.~\eqref{Eq:evolutionTrajectory}. Here we introduce multi-components auxiliary field in row $\bm{\phi}=(\phi_1,\ldots,\phi_K)$, with individual fields $\phi_k$ identified as the coherent-state eigenvalues of bosonic operator, $\hat{a}_k|\phi_k\rangle = \phi_k |\phi_k\rangle$. The reduced density operator is then obtained as
\begin{align}
\hat{\rho}_s(t)
=
\big\langle \bm{0} \big|
\mathcal{T}
\exp\!\left[
-i\int_0^t d\tau\,\mathcal{L}_{\rm sm}(\tau)
\right]
\hat{\rho}(0)
\end{align}
assuming a factorized initial state
$\hat{\rho}(0)=\hat{\rho}_s(0)\otimes\ket{\bm{0}}$. Noting that the total generator $\mathcal{L}_{\rm sm}$ acts as Liouvillian in reduced system space while as Hamiltonian in auxiliary mode space.

In the Schr\"odinger picture, the dynamics of the embedded system follows
\begin{align}\label{Eq:qdmessHilbertSpace}
\dot{\hat{\rho}}(t) &= -i[\hat{H}_s,\hat{\rho}(t)] - i\hat{\bm{a}}^\dagger \mathcal{E}\hat{\bm{a}}\,\hat{\rho}(t) \notag\\
&\quad -i[\hat{S}, \left(\bm{\kappa}^\dagger\hat{\bm{a}}
+\hat{\bm{a}}^\dagger\bm{\eta}\right)\hat{\rho}(t)] 
-i \{\hat{S},  i\hat{\bm{a}}^\dagger\bm{\eta}' \hat{\rho}(t) \}.
\end{align}
The pure-state representation discussed above is isomorphic to a $2K$-dimensional Hilbert-space representation, with the right operators mapped to $K$ additional Hilbert-space operators. A numerically optimized form of Eq.~\eqref{Eq:qdmessHilbertSpace} using diagonal $\bm{\mathcal{E}}$ and $C(t)$ and $C^*(t)$ instead of real and imaginary parts was given in~\cite{xu2022taming}.

\emph{Example~1.--} 
In the underdamped regime $\omega_0 > \gamma_0$, the correlator \eqref{Eq:ClassicalBrownianCorrelationFunction} at high temperature $\beta \zeta \ll 1$ limits ($\zeta=\sqrt{\omega_0^2-\gamma_0^2}$) gives 
\begin{align}
    C(t\ge 0) = \frac{c_0^2}{\beta\omega_0^2} \phi_q(t) + i\frac{c_0^2}{2\omega_0} \phi_p(t) \, .
\end{align}
The functions $\phi_q(t) = {\rm e}^{-\gamma_0 t}[\cos{\zeta t} + (\gamma_0/\zeta)\sin{\zeta t}]$ and $\phi_p(t) = -(\omega_0/\zeta)~ {\rm e}^{-\gamma_0 t}  \sin{\zeta t}$ identify them as the position and momentum response of a damped oscillator~\cite[Sec.~7]{risken84}, respectively. The correlator can be expressed in form of \eqref{Eq:RICorrelatorDecomposition} with $\bm{\kappa}^\dagger = [1\;0]$, and 
\begin{align}
 \bm{\mathcal{E}} = -i\begin{bmatrix}
     0 & \omega_0 \\ -\omega_0 & 2\gamma_0
 \end{bmatrix},\,
 \bm{\eta} = \begin{bmatrix}
     \cfrac{c_0^2}{\beta\omega_0^2} \\ 0
 \end{bmatrix},\,
  \bm{\eta}' = \begin{bmatrix}
     0 \\ \cfrac{c_0^2}{2\omega_0}
 \end{bmatrix}.
\end{align}
The dynamics in Schr\"odinger picture reads 
\begin{align}\label{Eq:heomHilbertClassical}
   \dot{\hat{\rho}}(t) &= -\mathcal{L}_{\rm sm} \hat{\rho}(t) \notag \\
   &=-i[\hat{H}_s, \hat\rho(t)] -(2\gamma_0\hat{b}^\dagger\hat{b} + \omega_0\hat{a}^\dagger\hat{b} -\omega_0 \hat{a}\hat{b}^\dagger ) \hat{\rho}(t)  \notag \\
   &\quad -i(\frac{c_0^2}{\beta\omega_0^2}\hat{a}^\dagger + \hat{a}) [\hat{S}, \hat{\rho}(t)] + \frac{c_0^2}{2\omega_0} \{\hat{S},\hat{b}^\dagger\hat{\rho}(t)\}. 
\end{align}
We introduce ADOs by projecting the total density operator onto an unnormalized auxiliary modes Fock basis, $\rho_{m,n}(t) = (m,n|\hat{\rho}(t)$, where the basis states are defined by $\hat b\,|m,n) = -m\,|m-1,n)$ and $\hat a\,|m,n) = |m,n-1)$. The resulting equations of motion for $\rho_{m,n}(t)$ reproduce the HEOM given by Eq.~(20) of Ref.~\cite{li2022low}.

\emph{Example~2.--} 
Implementing the following Bogoliubov transformation 
\begin{align}
\mathcal{B}
&=e^{-\tfrac{1}{2}\left(\hat a^{2}+\hat b^{2}\right)}\,
e^{\frac{1}{2}\ln\{\tfrac{\beta \omega_0^2}{2 c_0^2}\} \left(\hat a^\dagger \hat a+\hat b^\dagger \hat b\right)}\,
e^{i\pi\,\hat b^\dagger \hat b }
\end{align}
to Eq.~\eqref{Eq:heomHilbertClassical} yields the transformed density operator $\mathcal{B} \hat{\rho}(t)$ that again denoted by $\rho(t)$, leading to equation of motion formulated in the Namba-Keldysh space:
\begin{align}\label{Eq:BogoHEOM}
   \dot{\hat{\rho}}(t) &=-i[\hat{H}_s, \hat\rho(t)] -[2\gamma_0 (\hat{b}^\dagger\hat{b} - \hat{b}^2) - \omega_0\hat{a}^\dagger\hat{b} +\omega_0 \hat{a}\hat{b}^\dagger] \hat{\rho}(t)  \notag \\
   &\quad -i\sqrt{\frac{c_0^2}{2\beta\omega_0^2}} [\hat{S}, (\hat{a}^\dagger +\hat{a})\hat{\rho}(t)]  \notag \\
   &\quad + \sqrt{\frac{\beta c_0^2}{8}} \{\hat{S},(\hat{b} - \hat{b}^\dagger)\hat{\rho}(t)\} \,.
\end{align}
Representing the equation in the unnormalized Fock basis defined by $\hat{b}|m,n) = m|m-1,n)$ and $\hat{a}|m,n) = n|m,n-1)$, recovers the high-temperature limit of the HEOM given in Eq.~(56) of Ref.~\cite{li2022low}. This result is significant: for spectral densities where the conventional truncated HEOM suffers from numerical instability at long times \cite{li2022low}, the Bogoliubov-transformed formulation of Eq.~\eqref{Eq:BogoHEOM} restores stability, as demonstrated numerically in the same reference.

\emph{Example~3.--} The effective mode propagator can also be constructed by introducing the basis functions $\phi_x(t) = \exp{-i(\zeta-i\gamma_0)t}$ and $\phi_y(t) = \exp{i(\zeta+i\gamma)t}$. The correlator can be expressed in \eqref{Eq:RICorrelatorDecomposition} with $\bm{\kappa}^\dagger = [1\;1]$ and 
\begin{align}
\bm{\mathcal{E}} \!=
\begin{bmatrix}
\zeta\!-\!i\gamma_0 & 0 \\
0 & -\zeta\!-\!i\gamma_0
\end{bmatrix},
\bm{\eta} \!=\! 
\begin{bmatrix}
\dfrac{c_1 \!+\! c_2^\ast}{2} \\[2mm]
\dfrac{c_1^\ast \!+\! c_2}{2}
\end{bmatrix},
\bm{\eta}' \!=\! 
\begin{bmatrix}
\dfrac{c_1 \!-\! c_2^\ast}{2i} \\[2mm]
\dfrac{c_2 \!-\! c_1^\ast}{2i} 
\end{bmatrix}. \notag
\end{align}
The corresponding generator is
\begin{align}
    \mathcal{L}_{\rm sm} &= [\hat{H}_s,\cdot] + (\zeta - i\gamma_0)\hat{a}^\dagger\hat{a} -(\zeta+i\gamma_0)\hat{b}^\dagger\hat{b} \notag \\
    & + [\frac{c_1+c_2^\ast}{\sqrt{2}} \mathcal{S}_q +\frac{c_1-c_2^\ast}{\sqrt{2}}\mathcal{S}_c]\hat{a}^\dagger + \sqrt{2}\mathcal{S}_q\hat{a}  \notag \\
    &+ [\frac{c_2+c_1^\ast}{\sqrt{2}}\mathcal{S}_q + \frac{c_2-c_1^\ast}{\sqrt{2}} \mathcal{S}_c]\hat{b}^\dagger + \sqrt{2}\mathcal{S}_q \hat{b}\,.
\end{align}
This generator corresponding to left space vectorization of the generator in \eqref{Eq:qdmessDiagonal}.

\emph{Example~4.--} The effective mode propagator in phase space can be constructed by introducing the basis functions
$\phi_q(t)=e^{-\gamma t}\cos(\zeta t)$ and
$\phi_p(t)=-ie^{-\gamma t}\sin(\zeta t)$. The correlator can be expressed in \eqref{Eq:RICorrelatorDecomposition} with $\bm{\kappa}^\dagger = [1\;0]$ and 
\begin{align}
\bm{\mathcal{E}} \!=\!
\begin{bmatrix}
-i\gamma & \zeta \\
\zeta & -i\gamma
\end{bmatrix},
\bm{\eta} = 
\begin{bmatrix}
\Re(c_1+c_2)\\
\Im(c_1-c_2)
\end{bmatrix},
\bm{\eta}' = 
\begin{bmatrix}
\Im(c_1+c_2)\\
\Re(c_2-c_1)
\end{bmatrix}. \notag
\end{align}
The corresponding generator is
\begin{align}
    \mathcal{L}_{\rm sm} &= [\hat{H}_s,\cdot] -i\gamma(\hat{a}^\dagger\hat{a}+\hat{b}^\dagger\hat{b}) +\zeta(\hat{a}^\dagger\hat{b} + \hat{a}\hat{b}^\dagger) + \sqrt{2}\mathcal{S}_q \hat{a} \notag \\
    & + \sqrt{2}[\Re{c_1 + c_2} \mathcal{S}_q + i\Im{c_1+c_2}\mathcal{S}_c]\hat{a}^\dagger \notag \\
    &+ \sqrt{2}[\Im{c_1-c_2}\mathcal{S}_q - i\Re{c_1-c_2}\mathcal{S}_c]\hat{b}^\dagger \,.
\end{align}

\section{Discussion}
\label{sec:discussion}

By exploiting the symmetry of the double time integral together with the Hermiticity of the bath correlation function, the self-energy can be expressed in several equivalent structural forms. For example, Eq.~\eqref{Eq:GeneralSelfEnergy} contains both time-ordered and anti-time-ordered contributions, whereas Eq.~\eqref{Eq:SelfEnergyRealRI} involves only time-ordered terms. These distinct representations lead to different unravelings of the self-energy in Liouville space or Hilbert space.

When combined with different decompositions of the bath correlation functions, as partially discussed in Sec.~\ref{sec:hspir}, these structural forms admit a left–right factorization, as outlined in Sec.~\ref{sec:factorization}. In such cases, the associated Green's function $\mathbb{G}(t)$ acquires simplified matrix structures that can be unraveled using deterministic coherent-state fields. Representative examples include triangular block structures [\eref{Eq:kmatrix}], block-diagonal forms [\eref{Eq:diagonalGreenMatrix}], and the representation given in Eq.~\eqref{Eq:HilbertSelfEnergy}. These constructions give rise to time-local Liouvillian generators for the embedded system dynamics.

In the present framework, Bogoliubov transformations are naturally encoded in the left–right factorization through Eq.~\eqref{Eq:LinearTransformation}. The corresponding similarity transformation acting on $\mathbb{G}(t)$ induces transformations between unraveling fields and their associated operators. Such transformations map the embedded-system generator between different representation frames that have been developed across various theoretical frameworks and research communities. For example, the generators in Eqs.~\eqref{Eq:GeneralLindbladpseudomodeEquation} (with $n_\beta = 0$) and \eqref{Eq:qdmessDiagonal} are connected via a Bogoliubov transformation \cite{xu2023universal,xu2026simulating}. The thermofield transformation, as a particular realization of a Bogoliubov transformation, is implicitly embedded in the dynamical modeling of auxiliary modes. This correspondence has been discussed in detail in Appendix~C of Ref.~\cite{xu2026simulating}.

Bogoliubov transformations preserve the exact physical dynamics: in the full Liouville space, they correspond to similarity transformations of the Liouvillian and therefore leave its spectrum invariant. However, practical implementations especially under strong coupling require truncation of the bosonic auxiliary-mode basis, and crucially, this truncation projection does not commute with the transformations. As a result, the finite-dimensional Liouvillian becomes representation dependent, and the approximate eigenvalue spectrum obtained after truncation may vary significantly between different frames \cite{yan2020new,li2022low}. Consequently, different representations can exhibit substantially different numerical stability, convergence behavior, and spectral properties. This non-commutativity, rather than being a mere technical complication, provides a valuable degree of freedom: by selecting an appropriate transformation, one can systematically redistribute truncation errors, stabilize simulations, optimize computational performance, and even gain alternative physical insights into the dynamics of strongly driven or strongly coupled open quantum systems.

\begin{acknowledgments}
We gratefully acknowledge financial support from the IQST and the German Research Foundation (DFG) under AN336/12-1 (FOR2724), AN336/17-1 and 524764293, from the BMBF within the project QSolid, and from the State of Baden-W\"urttemberg through the network KQCBW. Additionally, we  acknowledge the provision of computational resources by the state of Baden-Württemberg through bwHPC
and the German Research Foundation (DFG) through grant no.\ INST 40/575-1 FUGG (JUSTUS 2 cluster).
\end{acknowledgments}

\bibliography{quantum}

\end{document}